\documentclass[prb,amsfonts,twocolumn]{revtex4}

\def\vec#1{{\bf #1}}
\def\subsubsubsection#1{\vspace*{10pt} \noindent $\bullet$ {\sl  #1} : \\}

\def\F{P}

\begin{document}

\title{Generalized Quantum Hall Projection Hamiltonians}

\author{Steven H. Simon}

\affiliation{Lucent Technologies, Bell Labs, Murray Hill, NJ,
07974}

\author{E. H. Rezayi}

\affiliation{Department of Physics, California State University,
 Los Angeles California 90032}

 \author{Nigel R. Cooper}

 \affiliation{Cavendish Laboratory, Madingley Road, Cambridge, CB3 0HE, United Kingdom}


\begin{abstract}

Certain well known quantum Hall states --- including the Laughlin
states, the Moore-Read Pfaffian, and the Read-Rezayi Parafermion
states ---  can be defined as the unique lowest degree symmetric
analytic function that vanishes as at least $p$ powers as some
number $(g+1)$ of particles approach the same point. Analogously,
these same quantum Hall states can be generated as the exact
highest density zero energy state of simple angular momentum
projection operators.  Following this theme we determine the
highest density zero energy state for many other values of $p$ and
$g$.
\end{abstract}

\date{August 15, 2006}

\maketitle

\section{Introduction}

For two dimensional electron systems in very high magnetic fields,
the kinetic energy becomes fully quenched, electrons become
restricted to the lowest Landau level (LLL), and the effective
Hamiltonian is reduced to the potential energy of the
electron-electron interaction\cite{Prange}.  While naive intuition
might suggest that a Hamiltonian with only a potential energy
would result in a crystalline ground state, the analytic structure
of the lowest Landau level puts enormous restrictions on the type
of wavefunctions that can exist.  It is this structure that is
responsible for all the richness of the fractional quantum Hall
effect.

In Laughlin's original explanation of the fractional quantum Hall
effect\cite{Prange}, he noticed that, due to the LLL analytic
structure, his trial state could be completely defined by stating
that the many particle wavefunction must vanish as a particular
power of the distance between two electrons. In particular, for
the Laughlin $\nu=1/m$ state, the wavefunction vanishes as $(z_1 -
z_2)^m$ as particles with position $z_1$ and $z_2$ approach each
other.  The highest density wavefunction with this property is
precisely the Laughlin state. It was discovered soon thereafter
that these Laughlin wavefunctions were in fact the exact unique
highest density zero-energy ground state of particles interacting
with particularly simple short range model
potentials\cite{Haldane,Trugman} that amount to projection
Hamiltonians. In this paper, we intend to focus on these two
related issues --- the manner in which wavefunctions vanish, and
the existence of simple model projection Hamiltonians.

To be more explicit, let us define $L_2$ to be the relative
angular momentum of two particles.  For electrons (which are
fermions), $L_2$ must always be odd and the minimum value of $L_2$
in the LLL is given by $L_2^{min} =1$. For bosons in a magnetic
field (or rotating bose condensates, which can be mapped to bosons
in a magnetic field\cite{Bosons}), $L_2$ must be even and
$L_2^{min}=0$. We can then define a projection operator
$P_{2}^{p}$ with $p$ to project out any state where any two
particles have relative angular momentum less than $L_2^{min} +
p$. In the Lowest Landau level, this projection operator is
precisely the above mentioned Hamiltonian that gives the Laughlin
$\nu=1/(L_2^{min} + p)$ state as its ground state when $p$ is
even.  In other words, this projection operator, when used as a
Hamiltonian, gives positive energy to any situation where the
wavefunction vanishes as $(z_1 - z_2)^m$ with $m < L^{min}_2 + p$,
leaving the Laughlin state as the unique highest density zero
energy (ground) state.  Note that for $p$ odd, the wavefunction
cannot vanish as $p$ powers, so $P_2^p$ has the same effect as
$P_2^{p+1}$ in that both forbid relative angular momentum of $p-1$
or less.

Another very interesting set of trial wavefunctions have also been
studied that follow very much in this spirit. The Read-Rezayi
$Z_g$ parafermionic wavefunctions\cite{ReadRezayi} are the unique
exact highest density  zero energy (ground) state of simple
$(g+1)$ body interactions. Correspondingly, these wavefunctions
can be completely defined by specifying the manner in which the
wavefunctions vanish as $g+1$ particles come to the same point.
The Moore-Read Pfaffian\cite{MooreRead} state, which is thought to
be the ground state wavefunction for the observed $\nu=5/2$
plateau\cite{Pfaff}, is  precisely the $g=2$ member of this
series. In addition, the particle hole conjugate of the $g=3$
Read-Rezayi state has been proposed to be a candidate for the
observed $\nu=12/5$ fractional quantum Hall state\cite{Xia}.
Finally, we note that the $g=1$ element of this series is just the
Laughlin state with $p = 1$ or $p=2$.

Analogously to our above construction for the Laughlin series, we
may define $L_{g+1}$ to be the relative angular momentum of a
cluster of $g+1$ particles. It can be shown (and we will show
below) that for electrons in the LLL, the minimal value of
$L_{g+1}$ is given by $L_{g+1}^{min}  = g(g+1)/2$. For bosons, the
minimal value would be $L_{g+1}^{min} =0$.
 Symmetry dictates (as shown in appendix \ref{sec:app1}) that $L_{g+1} =
L_{g+1}^{min} +1$ cannot occur, although any other value of
$L_{g+1} \geq L_{g+1}^{min}$ can occur for $g>1$  (and $L_2$ must
be even or odd for bosons or fermions respectively). Again we
define $P_{g+1}^p$ to be a projection operator that projects out
any state where any cluster of $g+1$ particles has relative
angular momentum $L_{g+1} < L_{g+1}^{min}+p$. The Read-Rezayi
state can then be obtained\cite{ReadRezayi} from using the
projection operator $P^2_{g+1}$ as a Hamiltonian in the lowest
Landau level. [ Note that since $L_{g+1} = L_{g+1}^{min} + 1$ is
not allowed, the effect of $P^1_{g+1}$ and $P^2_{g+1}$ are both
the same in that they give nonzero energy to states where any
cluster has relative angular momentum $L_{g+1}= L_{g+1}^{min}$ ].
In this work we will consider the obvious generalization of the
Read-Rezayi construction, taking the Hamiltonian in the LLL to be
given by the projection operator $P_{g+1}^p$ for general $g$ and
$p$.

The general restriction that the minimum relative angular momentum
of $g+1$ particles be $L_{g+1} \geq  L_{g+1}^{min}+ p$ can be
expressed in terms of how the wavefunction vanishes as $g+1$
particles approach each other. For bosons, where $L_{g+1}^{min}
=0$, the wavefunction does not need to vanish as $g$ particles
approach a given position $\tilde z$ but as the $g+1^{st}$
particle arrives, the wavefunction must vanish as $(z_{g+1}-\tilde
z)^p$. The situation for fermions, however, is a bit more
complicated, and will be discussed in section \ref{sec:jas1}
below.

The purpose of this paper is to determine the highest density zero
energy state of the proposed Hamiltonian $P_{g+1}^p$ which is a
natural generalization of the Laughlin, Moore-Read, Read-Rezayi,
Haffnian and Gaffnian Hamiltonians. While we will not find a
solution for arbitrary $g$ and $p$, we will be able to find a
solution for many values of $g$ and $p$ that have not been
previously discussed. We note that in addition to the Laughlin
states ($g=1$ with any $p$)  and the Read-Rezayi states ($p=1$ or
$p=2$ with any $g$), the ground state of the $g=2$ and $p=4$ case,
know as the ``Haffnian" has been previously discussed by
Green\cite{Green}. In addition, the ground state of $g=2$ and
$p=3$ has been dubbed the ``Gaffnian", and is discussed in depth
in a companion paper by the current authors\cite{Gaffnian}.   (The
name ``Gaffnian" is an alpha-phonetic interpolation between the
$p=2$ pFaffian and the $p=4$ Haffnian).

The outline of this paper is as follows. We begin by fixing
notations and conventions in section \ref{sub:prelim}.  In section
\ref{sec:General} we define the concept of a "proper" cluster
wavefunction which is crucial to our arguments.  Through much of
this paper we focus on boson wavefunctions.  In section
\ref{sec:res2} we start filling out a table as to the highest
density ground state of the Hamiltonian $P^p_{g+1}$.  Although we
do not fill in all possible values of $p$ and $g$, we do determine
quite a few (results are given in Table I).   In section
\ref{sec:jas1} we discuss attaching jastrow factors to the
resulting wavefunctions, and in particular the fermionic analogues
of these wavefunctions.  We find that the structure of the table
for fermions and bosons is identical.


\subsection{Preliminaries}
\label{sub:prelim}

We will always represent a particle's coordinate as an analytic
variable $z$. On the plane $z = x + iy$ is simply the complex
representation of the particle position $\vec r$.  On the sphere,
$z$ is the stereographic projection of the position on the sphere
of radius $R$ to the plane.  All distances will be measured in
units of the magnetic length.  In symmetric gauge, single particle
lowest Landau level wavefunctions $\varphi({\bf r})$ are given as
analytic functions $\psi(z)$ times a measure $\mu(\vec r)$.
\begin{equation}
    \varphi(\vec r) = \mu(\vec r) \psi(z)
\end{equation}
On the disk the measure is\cite{Prange}
\begin{equation}
    \mu(\vec r) = e^{-|z|^2/4}
\end{equation}
whereas on the sphere (with stereographic projection) the measure
is\cite{ReadRezayi}
\begin{equation}
    \mu(\vec r) = \frac{1}{[1 +  |z|^2 / (4 R^2)]^{1 + N_\phi/2}}
\end{equation}
with $N_{\phi}$ being the total number of flux penetrating the
sphere.  On the sphere the degree of the polynomial $\psi(z)$
ranges from $z^0$ to $z^{N_{\phi}}$ giving a complete basis of the
$N_{\phi} +1$ states of the LLL.  On the disk, the degree of
$\psi$ can be arbitrary.

We will write multiparticle wavefunctions $\Psi$ for $N$ particles
in the lowest Landau level as an analytic functions $\psi$ of $N$
variables times the measure $\mu$
\begin{equation}
    \Psi(\vec r_1, \ldots, \vec r_N) =     \psi(z_1, \ldots,
    z_N) \mu(\vec r_1, \ldots, \vec r_N)
\end{equation}
with
\begin{equation}
\label{eq:measure}
    \mu(\vec r_1, \ldots, \vec r_N) = \prod_{i=1}^N \mu(\vec r_i)
\end{equation}
On the sphere, the polynomial $\psi$ cannot be of degree greater
than $N_{\phi}$ in any variable $z_i$.

A quantum Hall ground state wavefunction will be a
translationally, rotationally invariant quantum liquid.   The
restriction we impose on $\psi$ is that it must be a
translationally invariant homogeneous polynomial of degree
$N_\phi$.  On the sphere, the degree $N_\phi$ is just the number
of flux through the sphere.  Conversely, given a (translationally
and rotationally invariant) quantum Hall wavefunction on a sphere,
the flux $N_\phi$ can be identified as the highest power of $z_i$
that occurs.  We note that so long as our interaction in the
lowest Landau level is time reversal invariant, we can (and will)
choose the the polynomial $\psi$ with real coefficients of all
terms.  As the size of a system is extrapolated to the
thermodynamic limit, we have the relation
\begin{equation}
    N_{\phi} = \frac{1}{\nu} N - {\cal S}
\end{equation}
with $\nu$ the filling fraction, and $\cal S$ is known as the
``shift".  We note that on a torus geometry there is typically no
shift\cite{HaldaneRezayiTorus}.

For a bosonic wavefunction $\psi$ must be symmetric in its
arguments, whereas for a fermionic wavefunction it must be
antisymmetric in its arguments.  A well known theorem tells us
that any antisymmetric function can be written as a single
Vandermonde determinant times a bosonic function.  In this way we
can generally write
\begin{equation}
\label{eq:bosontofermion} \psi_{\mbox{\tiny fermion}}(z_1, \ldots,
z_N) = J \psi_{\mbox{\tiny boson}}(z_1, \ldots, z_N)
\end{equation}
where
\begin{equation}
\label{eq:jastrow}
    J = \prod_{i < j} (z_i - z_j)
\end{equation}
Using this relation, the translation from bosons to fermions is
quite easy.  It is easy to see that the filling fraction $\nu_f$
for fermions is related to that of the corresponding filling
fraction for bosons $\nu_b$ via
\begin{equation}
\label{eq:bosenutofermionnu}
 \nu_f = \frac{\nu_b}{\nu_b + 1}
\end{equation}
Throughout much of this paper we will be focused on bosonic
wavefunctions for clarity.   We will return to the issue of
fermionic wavefunctions briefly in section \ref{sec:jas1} below.

\section{Proper Cluster Wavefunctions}

\label{sec:General}

We begin by focusing on bosonic wavefunctions. A $g$-cluster
wavefunction $\psi$ will be defined by the analytic manner in
which the wavefunction vanishes when the $g+1$ particles are
brought to the same point $\tilde z$. Generally, we will write
this $g+1$ particle limiting behavior
\begin{widetext}
\begin{equation}
\label{eq:cond1}
    \lim_{z_1, \ldots, z_{g+1} \rightarrow \tilde z} \psi(z_1, \ldots,
    z_N) \sim  f(z_1, \ldots, z_{g+1}) \tilde \psi(\tilde z ; z_{g+2} \ldots z_{N})
\end{equation}
\end{widetext}
where $f$ is assumed to be an overall symmetric, translationally
invariant, homogeneous polynomial of degree $p$ (By
translationally invariant, we mean that we must have $f$ invariant
under shifting all $z_i \rightarrow z_i + \alpha$).  The relative
angular momentum of such a $g+1$-cluster is defined to be
$L_{g+1}= p$ on the disk.  Thus, on the disc a group of $g+1$
particles is not allowed to have relative angular momentum less
than $p$.

On the sphere, the notation is somewhat more
complicated\cite{Haldane}.  Each single particle state in the LLL
has angular momentum $N_{\phi}/2$.   The total angular momentum of
$g+1$ bosons in the same single particle state is then $(g+1)
N_{\phi}/2$.  If the relative angular momentum of the cluster is
$p$ then the total angular momentum of the cluster is $(g+1)
N_{\phi}/2 - p$

On the torus, no simple concept of angular momentum exists.
Indeed, the only way to describe the analogue appears to be to
specify the number of powers with which the wavefunction vanishes
(I.e., simply $p$).  Thus specifying $p$ appears to be more
universal than speaking in terms of angular momentum.

We assume that $f$ vanishes when all $g+1$ of its arguments
coalesce at the same point.  If $f$ does not vanish when $g$
particles coalesce, we say we have a ``{\sl proper}" $g$-clustered
wavefunction. If $f$ does vanish when $g$ or fewer particles
coalesce, then we say we have an ``{\sl improper}" $g$-clustered
wavefunction.

%


In the proper case, the fact that $f$ is homogeneous,
translationally invariant of degree $p$, tells us that when $g$
particles are put at the point $\tilde z$ we will have $f$
vanishing as $z^p$ as the $g+1^{st}$ particle approaches.
\begin{equation}
    \lim_{z \rightarrow \tilde z} f(\tilde z, \tilde z, \ldots, \tilde z, z)
     \sim (\tilde z - z)^p
\end{equation}
The wavefunction $\psi$ must vanish in this manner as {\it any}
$g+1^{st}$ particle approaches.  We can thus write that
\begin{eqnarray}
\label{eq:recursion} \nonumber \psi(\tilde z, \tilde z,  \ldots,
\tilde z, z_{g+1}, z_{g+2}, \ldots, z_N) &\sim& \\ \left[
\prod_{i={g+1}}^N (\tilde z - z_i)^p \right] \tilde \psi_1(\tilde
z; z_{g+1} \ldots z_{N}) & &
\end{eqnarray}
where $\tilde \psi_1$ is a wavefunction satisfying Eq.
\ref{eq:cond1} for the remaining $N-g$ particles (and may have
some dependence on $\tilde z$ as well).

Using this recursion relation, it is easy to calculate the filling
fraction and shift of this wavefunction.  We claim that for a
proper $f$ of degree $p$ (i.e., one that does not vanish when $g$
of its arguments come to the same point), the densest wavefunction
satisfying condition \ref{eq:cond1} occurs at flux $N_{\phi} = p
N/g - p$ so long as $N$ is a multiple of $g$. Thus, this
wavefunction has filling fraction and shift
\begin{equation}
\label{eq:propernu} \nu = g/p  ~~~~~~~~~~~~ {\cal S} = p.
\end{equation}
To see this result more explicitly, we imagine bringing together
particles into groups of $g$ particles and using the above
recursion relation (Eq. \ref{eq:recursion}) a total of  $N/g - 1$
times.  Let us put the first cluster of particles at position
$\tilde z_1$, the second at position $\tilde z_2$ and so forth
until we have grouped the $N/g-1^{th}$ group at position $\tilde
z_{N/g-1}$.  The last $g$ particles we leave ungrouped.   Using
the recursion law we obtain a wavefunction
\begin{widetext}
\begin{eqnarray} \nonumber
& & \psi(\tilde z_1,  \ldots, \tilde z_1, \tilde z_2,
 \ldots \tilde z_2, \ldots, \tilde z_{N/g-1}, \ldots,
 \tilde z_{N/g-1}, z_{N-g}, z_{N-g+1}, \ldots z_N) = \\
 & & \prod_{1 \leq a < b  \leq N/g - 1} (\tilde z_a - \tilde
 z_b)^{pg}  \prod_{1 \leq i \leq N/g - 1}\,\,\,\,\,\, \prod_{k= N - g}^N
 (\tilde z_i - z_k)^{p} \,\,\, \chi_{N/g-1}(\tilde z_1, \ldots, \tilde z_{N/g-1}; z_{N-g}, z_{N-g+1}, \ldots z_N)
\end{eqnarray}
\end{widetext}
where $\chi_{N/g-1}$ is not allowed to vanish as any of its $g$
remaining arguments $z_j$ coalesce.  The highest density
wavefunction satisfying the limiting behavior Eq. \ref{eq:cond1}
(i.e., the quantum Hall state with no quasiholes) could thus have
$\chi$ being unity.    Examining the degree of this polynomial
with respect to the position of $z_N$ we see that it is of degree
$p (N/g - 1)$. Thus, we have a wavefunction corresponding to flux
$N_\phi = p (N/g - 1) = (p/g) N - p$ which indicates $\nu=g/p$ and
${\cal S}=p$ as claimed.

For each proper function $f$, there exists {\it at most} one
corresponding quantum Hall ground state wavefunction which would
be the maximum density translationally invariant wavefunction for
which Eq. \ref{eq:cond1} is always obeyed.  Of course, just
because we have constructed an appropriate $f$ for $g+1$
particles, it is not clear how one can construct a wavefunction
with a large number of particles $N$ such that Eq. \ref{eq:cond1}
is obeyed as any combination of $g+1$ particles approach each
other. In essence we are asking how to ``sew" together many
functions $f$ to form a macroscopic wavefunction. Sometimes no
such macroscopic wavefunction exists.   For example, in Appendix
\ref{sec:qhf} it is shown that for odd $pg$ no such macroscopic
wavefunction exists.  We note, however, that many proper cluster
wavefunctions are already known. The $Z_g$ Read-Rezayi states, for
example are proposer $p=2$ states for any $g$ (including the
Pfaffian, which is $g=2, p=2$). The Laughlin states are proper for
$g=1$ with even $p$. The Haffnian state\cite{Green} is proper with
$g=2$, $p=4$ case, and recently the current authors\cite{Gaffnian}
have studied the ``Gaffnian" state, which is proper with $g=2$,
$p=3$.  Further, in the next section we will not need to know that
any more proper wavefunctions actually exist.  What is important
is that {\it if} they exist, we know what their filling fractions
are.

\section{Main Results}
\label{sec:res2}

We now examine possible pair combinations of $g$ and $p$  and ask
what the ground state is of the projection Hamiltonian
$P_{g+1}^p$.   Again we will consider here only the case of
bosons.  These results are summarized in Table I. In many of the
examples below, we will use the same type of reasoning: A
wavefunction that vanishes as $g+1$ particles come together must
be either improper or proper (either it does or does not vanish as
only $g$ particles come together).  We determine the densest
possible zero energy state for both of the two possibilities and
then compare these two with each other to find the densest of all
possible zero energy states.

\subsubsubsection{$g=1$: the Laughlin series}

The Hamiltonian $P_2^p$ gives positive energy to any pair of
particles with relative angular momentum less than $p$.  This
leaves the highest density zero energy ground state being the
$\nu=1/p$ bosonic Laughlin state for even $p$.  For odd $p$, the
Hamiltonian does not allow pairs to have relative angular momentum
$p-1$ so the highest ground zero energy ground state is the
$1/(p+1)$ bosonic Laughlin state.

\subsubsubsection{$p=1, p=2$: the Read-Rezayi series}

As discussed in the introduction, it has been
shown\cite{ReadRezayi} that projecting out the minimal angular
momentum of $g+1$ particles  (projecting out $L_{g+1}=0$ for
bosons) results in the $Z_g$ Read-Rezayi state.   Since $L_{g+1}
\neq 0$ as shown in Appendix \ref{sec:app1}, we then conclude that
the highest density zero energy state of both $P_{g+1}^1$ and
$P_{g+1}^2$ is the $Z_g$ Read-Rezayi state whose filling fraction
is $\nu=g/2$ for bosons. Note that this includes $g=2$ with
$p=1,2$ which gives the Moore-Read state (which is just the $g=2$
member of the Read-Rezayi series).

\subsubsubsection{$g=2$, $p=3$ Gaffnian}

The case $g=2, p=3$ give the Gaffnian state\cite{Gaffnian}.  We
need not go into much detail as to the physics of this state but
to indicate that such a proper cluster wavefunction at $\nu=2/3$
for bosons exists. Detailed discussion of this wavefunction is
given in Ref. \onlinecite{Gaffnian}.  For completeness, we now
consider also the possibility that the ground state is not a
proper cluster wavefunction, but rather an ``improper"
wavefunction (meaning it vanishes as only two particles come
together).   However, we know that the highest density bosonic
wavefunction that vanishes when two come together is the Laughlin
$\nu=1/2$ state, which is not as dense as the Gaffnian.

\subsubsubsection{$g=2$, $p=4$ Haffnian}

Similarly, the $g=2, p=4$ case give the Haffnian\cite{Green}.
Again, this is a proper cluster wavefunction for $\nu=1/2$ for
bosons has been previously discussed in detail.   Again, we
consider the possibility that the highest density state is an
improper wavefunction.  Indeed, the highest density improper
wavefunction is the Laughlin $\nu=1/2$ state which which vanishes
even faster than the Haffnian as 3 particles come to the same
point (so it is also a zero energy state of $P_{3}^4$). Comparing
these two possibilities, the Haffnian is considered the ground
since it has a shift of ${\cal S}=4$ whereas the Laughlin
$\nu=1/2$ state has a shift of ${\cal S}=2$. Thus the filling
fraction of the Haffnian is slightly greater by an amount order
$1/N$ (with $N$ the number of particles).  Note, however, on a
torus geometry, where there is no shift, the density of these two
states is the same (and indeed, there are many other states with
the same density too\cite{Green,Edunpub}).

 \subsubsubsection{The $g=2$ series
for $p=5,6$}

Let us start by considering the cases of $g=2$ and $p=5,6$.
Suppose the highest density ground state is a proper cluster
wavefunction.    In this case, the filling fractions in these two
cases would be $\nu=2/5$ and $\nu=2/6$ respectively (See Eq.
\ref{eq:propernu}). We now consider the possibility that the
ground state is improper. The highest density improper state (ie,
state that vanishes as two particles come together) is the
Laughlin $\nu=1/2$ state. This is denser than the proper
possibilities. Furthermore the Laughlin $\nu=1/2$ state is also a
zero energy state of the relevant Hamiltonians $P^5_3$ and $P^6_3$
since the Laughlin state vanishes as 6 powers when three particles
come together.  Thus we conclude that the Laughlin $\nu=1/2$ state
is the densest zero energy state of these Hamiltonians.

\subsubsubsection{The Periodic $g=2$ series}

For $p>6$, we proceed similarly.  If the highest density ground
state is proper, the filling fraction is $\nu=g/p$. Now suppose
the ground state is improper.  In this case, the wavefunction must
vanish as two particles come together.  It is well known that any
symmetric wavefunction $\psi$ that vanishes as two particles come
together can be written as two Jastrow factors (See Eq.
\ref{eq:jastrow}) times another symmetric wavefunction $\psi'$
\begin{equation}
\psi(z_1, \ldots, z_N) = J^2 \psi'(z_1, \ldots, z_N)
\end{equation}
(Compare Eq. \ref{eq:bosontofermion}).  The filling fraction $\nu$
of $\psi$ is related to the filling fraction $\nu'$ of $\psi'$ via
\begin{equation}
\nu = \frac{\nu'}{2 + \nu'}
\end{equation}
This is analogous to the usual composite fermion transformation
(compare also Eq. \ref{eq:bosenutofermionnu}).   Further, if
$\psi$ vanishes as $p$ powers when $3$ particles come together,
then $\psi'$ vanishes as $p'=p - 6$ powers (the $6$ being from the
Jastrow factors). Thus, if $\psi$ is improper with $g=2$ we are
equivalently looking for a wavefunction $\psi'$ that vanishes at
least as $p-6$ powers when 3 particles come together. Thus, we
discover that the highest density improper wavefunction for
$6<p\leq 12$ is just two Jastrow factors times the ground state of
$P^{p-6}_2$.  For $p \leq 6$ we have already calculated the ground
state of $P^p_2$ (i.e., $p=1,2$ is Pfaffian, $p=3$ is Gaffnian,
$p=4$ is Haffnian, and $p=5,6$ is Laughlin), thus we know the
highest density improper ground state of $P^p_2$ for $6 < p \leq
12$. It is easy to verify that the filling fraction of this
improper state is greater than the $\nu=g/p$ proper possibility.
For $12 < p \leq 18$ we can repeat the argument and find that it
is again the same series but with 4 Jastrow factors and so forth.

\subsubsubsection{Read-Rezayi Series again for $p=3,4$}

We now consider the case of $p=3,4$ for arbitrary $g$.  If the
highest density state is a proper $g$-cluster wavefunction then
the filing fraction will be $\nu=g/p$ as usual.   If the
wavefunction is improper,  then it must vanish as only $g$
particles come together.    But we already know that the highest
density state that vanishes as $g$ particles come together is the
$Z_{g-1}$ Read-Rezayi state whose filling fraction is
$\nu=(g-1)/2$. Furthermore, as shown in Appendix \ref{app:RR} the
$Z_{g-1}$ Read-Rezayi wavefunction vanishes as $4$ powers when
$g+1$ particles come together (for $g>1$).  Thus, so long as
$(g-1)/2 > g/p$, the Read-Rezayi $Z_{g-1}$ state is the highest
density zero energy state of $P^3_{g+1}$ and $P^4_{g+1}$.   Note
that this inequality is satisfied for $g>2, p=4$ and $g>3, p=3$.

\subsubsubsection{The $g=3, p=3$ Pfaffian}

For the $g=3,p=3$ case, the above inequality ($(g-1)/2 > g/p$) is
instead an equality.  Thus, this case is marginal.  Here, the
putative proper state occurs at $\nu=1$, and the improper state is
the $Z_2$ Read-Rezayi state  (the Moore-Read Pfaffian) which is
also $\nu=1$. The shift of the Pfaffian is ${\cal S}=2$, where as
the shift of a $p=3$ proper state should be ${\cal S}=3$. Thus, we
would expect that the proper state is denser. However, in appendix
\ref{sec:qhf} we show that, by symmetry, no proper state can exist
for $pg$ odd as we have in this case.  So there is no wavefunction
at $\nu=1$ with shift ${\cal S}=3$. Thus, the Pfaffian is the
densest possible zero energy state of $P^3_3$.   In this case, we
do not eliminate the possibility that another zero energy state
may exist with exactly this filling fraction (and perhaps the same
shift).   An otherwise ``proper" state where a term has been added
to fix the symmetry could occur.  Indeed, exact diagonalization on
the torus has revealed at least one other zero energy state at the
same filling fraction.

\subsubsubsection{The $g=3, p=5,6$ States : Gaffnian Conjecture}

We again consider first the possibility that the ground state of
$P^5_4$ and $P^6_4$ are proper.  These wavefunctions would have
filling fractions $3/5$ and $3/6$ respectively.   The other
possibility is that the highest density ground state is improper
(ie, it vanishes as only three particles come together).  Now
consider the Gaffnian wavefunction.  This has filling fraction
$2/3$, and from the explicit form of the wavefunction given in
Ref. \onlinecite{Gaffnian} it can be seen that it vanishes as 6
powers when 4 particles come together. Hence, the highest density
ground states of $P^5_4$ and $P^6_4$ must be improper.  However,
there could be another (improper) zero energy state that also
vanishes as 3 particles come together which is higher density than
the Gaffnian. We conjecture that the Gaffnian is indeed the
highest density zero energy state in these cases. However, we have
not been able to prove this conjecture.

\begin{widetext}
\vspace*{10pt}
 \small \begin{tabular}{|c
||c|c|c|c|c|c|c|c|c|c|c|c|l }
  \hline
\small   & $p=1,2$ & $p=3$ & $p=4$ & $p=5$ & $p=6$ & $p=7$& $p=8$
& $p=9$ & $p=10$ & $p=11$ & $p=12$ & $p=13$ & \ldots
 \\
  \hline
  \hline
  $g=1\, \rule[-6pt]{0pt}{18pt}$ & ~ J${}^2: \frac{1}{2}$
  & ~ J${}^4: \frac{1}{4}$&
   ~ J${}^4: \frac{1}{4}$& ~ J${}^6: \frac{1}{6}$& ~ J${}^6: \frac{1}{6}$& ~ J${}^8: \frac{1}{8}$
   & ~J${}^8: \frac{1}{8}$&~  J${}^{10}: \frac{1}{10}$& ~ J${}^{10}:
\frac{1}{10}$& ~ J${}^{12}:
   \frac{1}{12}$
   & ~ J${}^{12}: \frac{1}{12}$    &  ~ J${}^{14}: \frac{1}{14}$
  \\
  \hline
    $g=2\, \rule[-6pt]{0pt}{18pt}$
& ~ \F $: 1$ & ~ G $ : \frac{2}{3}$ & ~ *H $: \frac{1}{2}$ & ~
J${}^2: \frac{1}{2}$  & ~ J${}^2: \frac{1}{2}$ & ~ \F J${}^2:
\frac{1}{3}$ & ~ \F J${}^2: \frac{1}{3}$ & ~ GJ${}^2: \frac{2}{7}$
& ~ *HJ${}^2: \frac{1}{4}$ & ~ J${}^4: \frac{1}{4}$  & ~ J${}^4:
\frac{1}{4}$  & ~ \F J${}^4: \frac{1}{5}$
\\
  \hline
  $g=3\, \rule[-6pt]{0pt}{18pt}$
& ~ R${}_3$ $: \frac{3}{2}$  & ~ *\F $: 1$ &  ~ \F $: 1$ & & & & &
& & & &
\\
\hline
  $g=4\, \rule[-6pt]{0pt}{18pt}$
& ~ R${}_4$ $: 2$ & ~ R${}_3$ $: \frac{3}{2}$  & ~ R${}_3$ $:
\frac{3}{2}$ & & & & & & & & &
%
%
\\
\hline
  $g=5\, \rule[-6pt]{0pt}{18pt}$
& ~ R${}_5$ $: \frac{5}{2}$ & ~ R${}_4$ $: 2$  & ~ R${}_4$ $: 2$ &
& & & & & & & &
\\
\hline
  $g=6\, \rule[-6pt]{0pt}{18pt}$
& ~ R${}_6$ $: 3$ & ~ R${}_5$ $: \frac{5}{2}$  & ~ R${}_5$ $:
\frac{5}{2}$
& & & & & & & & &
%
\\
\hline
  $\vdots \, \rule[-6pt]{0pt}{18pt}$ &
$ \vdots $ & $\vdots$  &  $\vdots $ & & & & & & & & &
\end{tabular}
\vspace*{10pt}

\begin{minipage}[t]{7in} Table I: Highest density zero energy ground state of bosons with Hamiltonian $P^p_{g+1}$.
The entries in this table are ``Name of state" followed by filling
fraction.  Abbreviations are P = Pfaffian; G= Gaffnian;
H=Haffnian; J${}^n$= Jastrow Factor to the $n^{th}$ power; R${}_n$
= Z${}_n$ Read-Rezayi state.  So for example, the $g=2$, $p=9$
slot has G J${}^2 \,: \, \frac{2}{7}$ which means the wavefunction
is the gaffnian times 2 Jastrow factors which occurs at filling
fraction 2/7.  Note that Laughlin states are listed only as
J${}^n$.   An asterisk indicates that the state is ``marginal" in
that there are other states competing with this state that differ
at most by a finite shift.    For fermions the structure of the
table would be  identical except that the filling fractions would
be related to these bosonic filling fractions by Eq.
\ref{eq:bosenutofermionnu}.
\end{minipage}

\vspace*{10pt}
\end{widetext}

\section{Adding Jastrow Factors}

\label{sec:jas1}

So far we have only considered bosonic wavefunctions. Given any
 bosonic wavefunction $\psi_0$ such as any of those discussed
above, we can construct wavefunctions
\begin{equation}
\label{eq:jas1}
    \psi = \psi_0 \prod_{i < j} (z_i - z_j)^M = J^M \psi_0
\end{equation}
For even $M$ this would then be another bosonic wavefunction,
whereas for odd $M$ this would be a fermionic wavefunction.  Of
particular interest is the $M=1$ case which was also discussed
above in Eq. \ref{eq:bosontofermion}.  Here, more generally, the
filling fraction $\nu$ of $\psi$ in terms of the filling fraction
$\nu_0$ of $\psi_0$ as
\begin{equation}
    \nu = \frac{\nu_0}{M + \nu_0}
\end{equation}

There is, of course, a one to one mapping between the possible
space of wavefunctions $\psi_0$ and those in the space of $\psi$.
The defining limiting behavior of the wavefunction $\psi$ is now
given by (Compare Eq. \ref{eq:cond1})
\begin{widetext}
\begin{equation}
    \lim_{z_1, \ldots, z_{g+1} \rightarrow \tilde z} \psi(z_1, \ldots,
    z_N) \sim  f(z_1, \ldots, z_{g+1}) \left[ \prod_{1 \leq i <j \leq g+1}
    (z_i - z_j)^M \right] \tilde \psi_0(\tilde z ; z_{g+2} \ldots z_{N})
\end{equation}
when $g+1$ particles come together and
\begin{equation}
    \lim_{z_1, \ldots, z_{k} \rightarrow \tilde z} \psi(z_1, \ldots,
    z_N) \sim  \left[ \prod_{1 \leq i <j \leq k}
    (z_i - z_j)^M \right] \tilde \psi_{0k}(\tilde z ; z_{k+1} \ldots z_{N})
\end{equation}
\end{widetext}
when $k < g+1$ particles come together.  In other words, the
wavefunction vanishes as the Jastrow factor only when less than
$g+1$ particles come together, and vanishes increasingly quickly
(as defined by the function $f$) when $g+1$ come together.  Thus,
if $f$ vanishes as $p$ powers when $g+1$ particles come together,
the wavefunction $\psi$ vanishes as $M g(g+1)/2 +  p$ powers when
$g+1$ particles come together.

Enforcing the presence of Jastrow factors is a well known
procedure. For bosons, $M=2$ is obtained by forbidding any two
particles to have relative angular momentum of zero.  In other
words, adding a term $P_2^2$ to the Hamiltonian will assure that
any zero energy wavefunction contains an overall $M=2$ Jastrow
factor. This term, $P^2_2$ is usually known as a $V_0$
interaction\cite{Haldane} since it projects out pairs of particles
with relative angular momentum zero. Similarly, to enforce an
$M=4$ Jastrow factor, one adds $P_2^4$ to the Hamiltonian (In the
usually nomenclature this is a $V_0$ term and a  $V_2$ term).  So,
for example, if a given wavefunction $\psi_0$ is the highest
density zero energy ground state of $P^p_{g+1}$ then $\psi = J^M
\psi_0$ should be the highest density zero energy ground state of
\begin{equation}
     P_2^M + P^{M g(g+1)/2 + p}_{g+1}
\end{equation}
with $M$ even.  It is interesting to note that in cases listed in
table I above, the term enforcing the Jastrow factors is not
needed. For example, the highest density zero energy state of
$P_3^3$ is the Gaffnian.  Thus, choosing any even $M$ we would
expect that the highest density zero energy state of $P_2^M + P^{3
M + p}_3$ should be $J^M$ times the Gaffnian.  It is interesting,
that in this particular case the highest density zero energy state
of $P^{3 M + p}_3$ is already $J^M$ times the Gaffnian without
including the Jastrow forcing term $P_2^M$.  This is an intreguing
phenomenon, and we do not know if it is general.

We now return to the case of Fermions.  As mentioned above in the
introduction (See Eq. \ref{eq:bosontofermion}), any fermi
wavefunction can be written as a bose wavefunction times a single
Jastrow factor.  Thus, by simply using a system of Fermions, an
$M=1$ Jastrow factor is automatically obtained.  We also note that
this immediately tells us that the minimum angular momentum of
$g+1$ fermions in the LLL is
\begin{equation}
    L^{min; fermion}_{g+1} = g(g+1)/2
\end{equation}
Since we have defined $P^p_{g+1}$ to project out relative angular
momenta $L < L^{min}+p$, the table generated as the highest
density zero energy state of $P^p_{g+1}$ is the same for fermions
as it is for bosons only the resulting fermion wavefunctions have
an overall Jastrow factor attached ($M=1$).

To add further Jastrow factors to a Fermionic wavefunction, we
follow the analogous scheme to the Bosonic case, projecting out
any pairs of fermions with the minimal angular momenta.  Thus, for
Fermions, our operator $P_2^2$ is defined to project out any pair
with minimum angular momentum less than $L=L_2^{min;
fermion}+2=3$. Thus, a zero energy state of $P_2^2$ for fermions
must have at least $M=3$ Jastrow factors in the wavefunction.
Conventionally such a term is known as a $V_1$ term of the
Hamiltonian. Similarly, a zero energy state of $P_2^4$ for
fermions must have at least $M=5$ Jastrow factors in the
wavefunction.   Note that, by construction, this again follows the
rule that the resulting wavefunctions will always be the bosonic
analogue times a single Jastrow factor.

\section{Discussion}

\label{sec:exp}

The wavefunctions we have constructed in this paper all stem from
reasonably simple Hamiltonians, which involve projecting out
clusters of particles with given angular momenta.  The simplicity
of this construction, is, of course, much of the attraction of our
theory.  It is interesting that the only fundamentally ``new"
state that has appeared on our table of states so far is the
Gaffnian, which will be discussed in depth in a companion to this
paper\cite{Gaffnian}. It would be interesting to fill in the rest
of Table I to see if any other new states might appear.

Some of the states that fit in our scheme are of course well known
and well established to occur in nature. For example, the Laughlin
states are certainly seen in the Lowest Landau level\cite{Prange}.
Also among the states that fit in our construction is the
Moore-Read Pfaffian\cite{MooreRead}, which is strongly though  to
be the explanation of the plateau seen in the first excited Landau
level\cite{Pfaff} at $\nu=2 + 1/2$.  In addition, there are
several states in our scheme that seem likely to be seen in
nature, although there remains some level of uncertainty. For
example, There is some evidence\cite{ReadRezayi} that the the
particle-hole conjugate of the $g=3$ Read-Rezayi state is a good
trial state for $\nu=2 + 2/5$, which has been observed
recently\cite{Xia}. A detailed discussion of the Gaffnian
wavefunction is given in a companion to this paper\cite{Gaffnian}.
Although the Gaffnian has extremely high overlap with $\nu=2/5$
there is reason to believe that the Gaffnian is a critical state
rather than a phase.

It is interesting to note that in the Lowest landau level, most of
the known physics appears to be outside of the general scheme set
out in this paper.   Instead, it appears that most of the states
seen in the LLL are most easily explained within a composite
fermion theory\cite{Olle}.  In contrast to the current work, the
composite fermion wavefunctions (with the exception of the
Laughlin states) are not the exact ground state of any known
simple Hamiltonian -- even though they are extremely accurate
wavefunctions for Coulomb (and similar) interactions in the LLL.
There are also possibilities that some of these states might be
observed in systems of cold atoms.  Rotating Bose condensates can
be thought of as Bosons in a magentic field and thus (if
sufficiently two dimensional) become quantum Hall systems
\cite{Bosons}.  In cold atom systems, experimentalists have been
extremely clever about designing Hamiltonians to have desired
interactions.  Indeed, a scheme has been devised\cite{CooperExact}
which essentially generate exactly the type of $g+1$-particle
interaction necessary to yield the Read-Rezayi cluster series.
Another approach to generating the Pfaffian in cold atoms have
also been proposed\cite{Gurarie} which does not rely on rotation.

Since the Hamiltonians we are proposing in this paper are
relatively simple, we might hope that clever experimentalists will
be able to devise systems in which these Hamiltonians are
realized.

{\bf Acknowledgements:}  EHR acknowledges support from DOE under
contract DE-FG03-02ER45981 (has this changed?).  The authors
acknowledge conversations with F. D. M. Haldane, N. Read, and I.
Berdnikov.

\appendix

\appendix
\section{$L_{g+1}\neq L_{g+1}^{min}+1$}
\label{sec:app1}

The statement that $g+1$ bosons have relative angular momentum $p$
is equivalent to saying that as the particles all approach the
same point, the wavefunction vanishes as a $p^{th}$ degree
polynomial $f$ in the sense of Eq. \ref{eq:cond1}. The function
$f$ must be a translationally invariant symmetric polynomial.  We
claim that no such polynomial exists of degree one.  To see this
we note that there is only a single symmetric polynomial in $g+1$
variable of degree one
\begin{equation}
    \sum_{i=1}^{g+1} z_i
\end{equation}
and under translation $z_i \rightarrow z_i + a$ this is not
invariant.  Thus we conclude that $g+1$ bosons cannot have
relative angular momentum $1$.  Writing any fermion wavefunction
as an overall Jastrow factor times a boson wavefunction (See Eq.
\ref{eq:bosontofermion}) one can then show that generally
$L_{g+1}$ cannot be $L_{g+1}^{min} +1$.

\section{Odd $pg$ proper boson wavefunctions do not exist}
\label{sec:qhf}

Here, we claim that when both $g$ and $p$ are odd no macroscopic
bosonic wavefunction exists with shift of $p$ for that $g$ and
$p$. To see this, we use the recursion relation Eq.
\ref{eq:recursion} (which is true as long as the wavefunction does
not vanish as $g$ particles coalesce, ie, as long as it is proper)
and group the particles into groups of $g$ at positions $\tilde
z_i$. The wavefunction of the clustered super-particles is given
by
\begin{equation}
\label{eq:super} \psi = \prod_{i < j} (\tilde z_i - \tilde
z_j)^{pg}
\end{equation}
However, a cluster of $g$-bosons must remain a bosonic object (ie,
the wavefunction is symmetric under interchange), whereas $pg$ is
odd. This tells us immediately that no such wavefunction can
exist.

\section{The Read-Rezayi Wavefunction}
\label{app:RR}

As shown by Ref. \onlinecite{Cappelli}, the bosonic Read-Rezayi
wavefunction can be written by dividing the particles into $g$
groups, giving Jastrow factors only between particles in the same
group, and then symmetrizing over all choices of which particle is
in which group. We will assume the total number of particles $N$
is divisible by $g$ and define the first group to be particles $1
\ldots N/g$ the second group to be $N/g + 1. \ldots 2 N/g$ and so
forth.  We thus write the $Z_g$ Read-Rezayi bosonic wavefunction
as
\begin{widetext}
\begin{equation}\label{eq:paraf}
    \psi = S_N\left[ \prod_{0 < i_1 < j_1 \leq N/g}
     (z_{i_1} - z_{j_1})^{2}   \prod_{N/g < i_2 < j_2 \leq 2N/g}
     (z_{i_2} - z_{j_2})^{2}  \,\, \ldots \,\,
     \prod_{(g-1)N/g < i_g < j_g \leq N}  (z_{i_g} - z_{j_g})^{2}
     \right]
\end{equation}
\end{widetext}
where $S_N$ represents symmetrization over all particle
coordinates.  It is trivial to establish that the filling fraction
is $\nu=g/2$ and the shift is ${\cal S}=2$.   When $g$ bosons come
together, one can be in each group so the wavefunction does not
vanish.  When $g+1$ bosons come together, at least two of them
must be in the same group and the wavefunction vanishes as $p=2$
powers. Similarly when $g+2$ particles come together (for $g>1$),
at least two groups have two bosons in them, meaning the
wavefunction vanishes as $p=4$ powers.


\begin{thebibliography}{22}

\bibitem{Prange} For a classic review of quantum Hall physics,
see R. Prange and S. M. Girvin eds, {\it The Quantum Hall Effect},
Springer-Verlag, NY (1987).

\bibitem{Haldane} F. D. M. Haldane, Phys. Rev. Lett. {\bf 51}, 605
(1983).

\bibitem{Trugman} S. Trugman and S. Kivelson, Phys. Rev. B{\bf
31}, 5280 (1985).

\bibitem{Bosons} See, for example, N. R. Cooper, N. K. Wilkin and J. M. F. Gunn,
Phys. Rev. Lett. {\bf 87}, 120405 (2001)

\bibitem{ReadRezayi}  N. Read and E. Rezayi, Phys. Rev.  {\bf B59}
8084 (1999).

\bibitem{MooreRead} G. Moore and N. Read, Nucl. Phys. {\bf B360} 362
(1991).

\bibitem{Pfaff} R. H. Morf, Phys. Rev. Lett. {\bf 80}, 1505
(1998); E. H. Rezayi and F. D. M. Haldane, Phys. Rev. Lett. {\bf
84}, 4685 (2000).

\bibitem{Xia} J. S. Xia et al, Phys. Rev. Lett. {\bf 93}, 176809
(2004).


\bibitem{Green} Dmitri Green, PhD Thesis, Yale University 2001; see cond-mat/0202455


\bibitem{Gaffnian} S. H. Simon, E. H. Rezayi, and N. R. Cooper,
companion paper.

\bibitem{HaldaneRezayiTorus} F. D. M. Haldane and E. H. Rezayi,
Phys. Rev. {\bf B}31, 2529 (1985).




%
%

\bibitem{Olle}   See
``Composite Fermions'', ed O. Heinonen, World Scientific, 1998;
and  therein.


\bibitem{Edunpub} E. H. Rezayi, unpublished.

\bibitem{Cappelli}  This is particularly clear using the form of the Read-Rezayi wavefunction written down by
A. Cappelli, L. S. Georgiev, and I. T. Todorov, Nucl Phys. {\bf
B599}, 499 (2001).
%
\bibitem{CooperExact} N. R. Cooper
Phys. Rev. Lett. 92, 220405 (2004).
%
\bibitem{Gurarie}   V. Gurarie, L. Radzihovsky, and A. V. Andreev,
cond-mat/0410620.
%


\bibitem{ReadRezayiPfaff} N. Read and E. H. Rezayi, Phys. Rev. {\bf B54}, 16864
(1996).

%



\end{thebibliography}
\end{document}